\RequirePackage{amsmath}
\documentclass[runningheads]{llncs}

\usepackage{graphicx}
\usepackage{amsmath}
\usepackage{amssymb}
\usepackage{hyperref}
\usepackage{authblk}

\title{Themed Challenges to Solve Data Scarcity in Africa: A Proposition for Increasing Local Data Collection and Integration}
\titlerunning{Themed Challenges to Solve Data Scarcity in Africa}

\author{Mubaraq Yakubu\inst{1, 2} \and Udunna Anazodo\inst{2} \and Maruf Adewole\inst{3} \and Theodore Barfoot\inst{1} \and Tiarna Lee\inst{1} \and Tom Vercauteren\inst{1} \and Jonathan Shapey\inst{1} \and Andrew King\inst{1} \and Alexander Hammers\inst{1}}

\institute{King's College London, London, United Kingdom \email{\{mubaraq.yakubu,alexander.hammers\}@kcl.ac.uk} \and
McGill University, 3801 Rue University, Montreal, Quebec, H3A 2B4, Canada \email{udunna.anazodo@mcgill.ca} \and
Medical Artificial Intelligence Lab, Lagos, Nigeria \email{madewole@mailab.io}}

\date{}

\begin{document}

\maketitle

\begin{abstract}
In Africa, the scarcity of computational resources and datasets stand as significant hurdles impeding the development and deployment of artificial intelligence (AI) tools in clinical environments and contributes to the increase in global bias. These constraints complicate the realization of AI’s full potential, thereby posing significant obstacles to the advancement of healthcare in the region. 
This paper presents a proposed framework to address data scarcity in healthcare across the continent. The framework outlines a comprehensive strategy to encourage healthcare providers in Africa to create, curate, and share local medical imaging datasets. Through themed challenges, designed to promote participation, accurate and relevant curated datasets within the African healthcare community may be created. By facilitating this approach, the existing dataset scarcity may be overcome, fostering a more inclusive and impactful AI ecosystem tailored to African healthcare needs.
\end{abstract}




\section{Introduction}
\label{sec:introduction}
In the landscape of African healthcare, the dearth of computational infrastructure and representative datasets has posed formidable challenges for the development of artificial intelligence (AI)  technologies. 
On the other hand, AI is transforming medical research and patient care through its diverse applications across various fields \cite{d2022artificial}. Emerging from high-income countries, there are large and carefully-curated datasets like the UK Biobank \cite{littlejohns2019uk}, The Cancer Imaging Archive (TCIA) \cite{clark2013cancer}  and the Alzheimer's Disease Neuroimaging Initiative (ADNI) \cite{petersen2010alzheimer}. 
However, Africa's contribution to publicly available neuroimaging datasets remains disproportionately low compared to developed countries or even other middle-income countries\cite{omotayo2024survey,wang2019statistical}.

Deficits in infrastructure, imaging facilities nearly exclusively operated by the private sector (clinics) rather than hospitals or universities, unreliable electricity supply, lack of centralised secure image data storage such as Picture Archiving and Communcation Systems (PACS), limited specialist availability, and a lack of knowledge and resources have compounded the obstacles hindering the growth and sustainability of healthcare initiatives across Africa \cite{geethanath2019accessible,ogbole2018survey,kiemde2020challenges,arakpogun2021artificial,anazodo2023framework}. 
Africa has the highest genetic diversity globally, encompassing more than 3,000 distinct ethnic groups~\cite{asiedu2024case}, so the absence of representative datasets for these ethnicities clearly limits the efficacy of AI tools in Africa, thwarting efforts to implement computer-assisted diagnosis and intervention effectively \cite{gwagwa2021responsible,kiemde2020challenges}. The data gap not only stymies the development of AI-driven healthcare solutions but also perpetuates disparities in healthcare outcomes, widening the chasm of global health inequities \cite{arakpogun2021artificial}. Racial bias has been shown even for seemingly innocuous tasks like brain segmentation \cite{ioannou2022study}. Addressing this issue is important for extending the transformative potential of AI to African healthcare.

Open challenges in medical image analysis have been instrumental in the advancement of machine learning algorithms \cite{mehta2022qu}. Drawing from the success of similar challenges, here we propose adopting this approach to gather local datasets that represent the African continent as part of themed challenges for specific brain diseaese(s), incentivising contributors to submit the best-curated datasets possible. By making data accessibility for others a prerequisite for participation, we aim to create a virtuous cycle through which data availability can grow. 
This endeavor can pave the way for dedicated AI tools tailored to local contexts. Equipping clinicians with AI tools trained on local datasets and trusted data can accelerate the adoption of computer-assisted diagnosis and intervention \cite{morley2020ethics}, driving advancements in African healthcare.

\section{Current Approaches and Efforts in Africa}
\label{sec:approaches}
Efforts to address healthcare challenges in Africa are already underway through a range of innovative initiatives and collaborative endeavors. We first review these initiatives before highlighting challenges and potential solutions in the African context.

\subsection{Open Data Initiative}

The Open Data initiative \cite{opendata2024webpage}, which will be launched at the Medical Imaging Computing and Computer Assisted Intervention (MICCAI) conference 2024, aims to foster collaboration and innovation in medical imaging research by promoting the sharing of diverse and inclusive datasets. With a focus on underrepresented populations and diseases, particularly from the African continent, the initiative seeks to address disparities in healthcare and promote inclusivity in AI research. Central to this initiative is the establishment of the AFRICAI repository, which will host publicly available high-quality medical imaging datasets. By encouraging researchers and data custodians to contribute datasets to the repository and adhering to FAIR \cite{morley2020ethics} data principles, the initiative aims to facilitate collaboration, reproducibility, and awareness of the importance of inclusivity in developing robust healthcare solutions.

\subsection{African Neuroimaging Archive (AfNiA)}

AfNiA is an initiative aimed at democratizing access to medical imaging data in Africa \cite{anazodomiccai}. By aggregating brain MRI scans and clinical attributes from multiple clinics across the continent, AfNiA facilitates the development and deployment of AI imaging innovations. Prioritizing ethical data use and governance, AfNiA ensures that AI solutions directly benefit the local communities that contribute data. 

\subsection{SPrint AI training for AfRican medical imaging Knowledge translation (SPARK) Academy}

SPARK Academy trains a new generation of African AI experts in medical imaging, bridging the gap between AI knowledge and clinical application (\cite{anazodobuilding}). It is an initiative from the Consortium for Advancement of MRI Education and Research in Africa (CAMERA) which is focused on improving (or sometimes even establishing) MRI education in Africa.

\subsection{Brain Tumor Segmentation (BraTS) Challenge 2023: Glioma Segmentation in Sub-Saharan Africa Patient Population (BraTS-Africa)}

The BraTS-Africa Challenge addresses the difficulties in diagnosing and treating gliomas in Sub-Saharan Africa (SSA) \cite{adewole2023brain}. Gliomas, the most common type of primary brain tumors \cite{reynoso2021epidemiology}, pose significant diagnostic and treatment challenges due to their aggressiveness and resistance to conventional therapy.
Late presentation of the disease at advanced stages and the use of lower-quality MRI technology exacerbate diagnostic difficulties in SSA populations. The BraTS-Africa Challenge seeks to develop and evaluate computer-aided diagnostic (CAD) methods tailored to resource-limited settings, where the potential for transformative healthcare impact of AI is highest by mitigating the workforce limitations.


\section{Difficulties Around AI Development in Healthcare in Africa and Potential Solutions}

\subsection{Resource Limitation}
Many healthcare institutions in Africa operate with constrained resources, facing limited access to advanced computational infrastructure like GPUs and high-speed internet, which hinders the implementation of AI solutions. This scarcity affects data processing capabilities and restricts the development of sophisticated AI models. Additionally, the socio-economic disparities in low- and middle-income countries (LMICs) result in insufficient investments in healthcare and research. The high number of low-income countries (LICs) and LMICs in Africa means government funding for AI health research is often inadequate \cite{nabyonga2021state}. This lack of investment hinders the acquisition of essential clinical data-generating devices, such as imaging equipment. Even when these devices are obtained, maintaining them—like MRI scanners—becomes challenging due to frequent breakdowns caused by intermittent electricity supply \cite{anazodo2023framework}.

Addressing these resource limitations requires increased funding and investment in basic infrastructure, healthcare, and research infrastructure. International collaborations, public-private partnerships, and initiatives aimed at enhancing technological capabilities in LMICs can play a pivotal role in overcoming these challenges. Additionally, providing avenues for local datasets to be shared with institutions that have better computational resources for training models could prove effective in mitigating these issues.

\subsection{Technical Skills Gap}
There is a significant shortage of healthcare professionals with AI expertise in Africa. This skills gap encompasses data management, annotation, and analysis, all of which are crucial for developing and maintaining robust AI systems.
Although efforts such as training programs, workshops, and academic courses have been initiated to address this gap \cite{anazodo2023framework,anazodobuilding,njamnshi2023brain,adewole2023brain}, there is still more to be done.

Expanding access to high-quality education and continuous professional development is essential to upscale the number of required skilled professionals. Creating supportive learning environments, such as AI research centers, and fostering collaborations between academia and industry, can enhance innovation. By scaling up these initiatives, a larger, more skilled workforce can be developed, driving AI advancements in African healthcare and improving health outcomes.

\subsection{Local Regulations and Bureaucracy}
Successfully navigating the local regulatory and bureaucratic landscape is crucial for the implementation of data collection and AI initiatives. Bureaucratic procedures can vary widely between regions and institutions \cite{koumamba2021health}, often involving complex approval processes that can delay or obstruct project progress. Understanding and effectively engaging with these procedures is essential for gaining necessary permissions and support.

\subsection{Building Trust and Overcoming Inaccurate Perceptions}
External researchers often face difficulties in building trust and gaining acceptance from local institutions and communities. There is a need to demonstrate genuine motives and commitment to local needs to foster collaboration. Additionally, local stakeholders may inaccurately perceive external researchers as sources of substantial funding, which can lead to unrealistic financial expectations and potential conflicts. Managing these perceptions and establishing transparent, mutually beneficial partnerships is vital for the success of data collection efforts.

\subsection{Regulatory and Ethical Issues}
Navigating the regulatory landscape across African countries poses significant challenges, particularly concerning ethical considerations in medical data use. As first outlined by Beauchamp and Childress, the four principles of biomedical ethics; namely respect for autonomy, beneficence, non-maleficence, and justice,provide a crucial framework for addressing these challenges \cite{woodman2019introduction}.
By prioritizing ethical standards, stakeholders can foster trust and responsibility in integrating AI into African healthcare, preventing the exacerbation of existing health disparities and promoting equitable distribution of AI advancements.

\subsection{Connectivity Issues}
Reliable internet connectivity remains a significant challenge in many parts of Africa \cite{dzinamarira2020covid}. Inconsistent or slow internet connections impede the ability to share and access data, collaborate with other researchers, and deploy cloud-based AI solutions. Addressing connectivity issues by introducing strategies that utilize offline tools, such as local data servers, edge computing, or portable data storage devices, is critical for the successful implementation of renowned effective AI approaches like federated learning and other collaborative approaches. 

\subsection{Data Scarcity}
The availability of well-curated medical image data in Africa is severely limited, hindering the training and validation of AI models which require large datasets for accuracy and reliability \cite{liang2022advances}. Local databases are often fragmented and lack standardization \cite{koumamba2021health}, making data aggregation challenging. Despite Africa's significant diversity, there is absence of vital data that represents the population \cite{bray2022cancer}, posing barriers to developing effective and generalizable AI tools \cite{kiemde2020challenges}.

Addressing data scarcity requires encouraging local data collection and integration initiatives. Establishing standardized protocols for data collection, curation, and sharing is crucial to building cohesive and comprehensive medical image databases. Such efforts would enhance the quantity and quality of available data, ensuring AI tools are relevant and beneficial to Africa's diverse population.

\section{Strategic Framework for Increasing Local Data Collection and Integration}

In response to the critical challenge of  of local and representative data  in African healthcare, we propose a strategic framework aimed at organizing and curating medical images by African clinicians and making the data publicly available. Inspired by the success of initiatives like the BraTS challenge, our strategy as described in the following paragraphs, aims to incentivize participation, ensure data quality, promote diversity, and foster sustainable practices in data collection and curation through themed challenges.

\subsection{Motivation and Incentivization}

To inspire and incentivize healthcare practitioners to establish local databases, our strategic framework outlines several key factors that we think will be addressed by organising the proposed themed challenges as follows:
\begin{enumerate}
    \item \textbf{Incentivizing Participation Through Themed Challenges:} Central to this strategy is the introduction of themed challenges within the African healthcare community. These challenges will task participants with collecting and curating datasets related to particular brain disease(s). To encourage participation and excellence, prizes will be awarded to participants with the best submitted curated datasets (e.g. best demographic descriptors, best validations of annotations), fostering a competitive yet collaborative environment for advancing medical imaging research (see also 4.2).
    
    \item \textbf{Ensuring Data Quality:} Recognizing the paramount importance of data quality, participation in the proposed challenge would be contingent upon meeting predefined criteria, which encompass aspects such as the development of standardized guidelines for data quality control, publishing annotation protocols \cite{radsch2023labelling}, data completeness, accuracy, and relevance to the target population. The themed challenges will thus establish stringent yet practical eligibility criteria and uphold the highest standards.

    \item \textbf{Adherence to Privacy and Ethics}: The strategy emphasizes the importance of data privacy and ethics. Clinicians would need to provide evidence of compliance with regulations such as General Data Protection Regulation (GDPR) and Health Insurance Portability and Accountability Act (HIPAA) before participating, including appropriate anonymisation and removal of patient-identifiable data. Additionally, people submitting data would need to evidence approval from a relevant overseeing organisation, typically an ethical review board to ensure appropriate consent for data sharing has been obtained, hence upholding ethical principles throughout the data collection and curation process.

    \item \textbf{Promoting Diversity and Representativeness}: Recognizing the importance of addressing data bias, one criterion for success in winning themed challenges will be consideration of dataset diversity. Clinicians would be encouraged to submit annotated datasets so that eventually the full spectrum of demographic, geographic, ethnic, and clinical variability within the African population is represented. This approach aims to mitigate the risk of bias and ensure the generalizability of AI models trained on the curated datasets.

    \item \textbf{Sustainability and Partnerships}: To ensure the sustainability of the strategy, we propose forming strategic partnerships with initiatives like AfNiA \cite{anazodo2023framework}, Open Data Initiative \cite{opendata2024webpage} and securing sponsorship opportunities with reputable organizations such as MICCAI, International Society for Magnetic Resonance in Medicine (ISMRM) Africa, and the African Society of Radiology. Leveraging the support and resources of these organizations will establish a long-term prospect that facilitates ongoing funding, human resources, and maintenance to sustain the strategy beyond its initial phase.
\end{enumerate}

\subsection{Motivation for Health Practitioners}
Healthcare practitioners are motivated to participate in the proposed strategy by several key benefits:

\begin{itemize}
    \item \textbf{Improved Patient Care}: Access by AI tool developers to local datasets enables the development of AI models tailored to specific patient needs, leading to more accurate diagnoses and personalized treatments.
    \item \textbf{Enhanced Research Opportunities}: Practitioners can engage in valuable research, contribute to medical advancements, and gain recognition as co-authors in collaborative studies.
    \item \textbf{Professional Development}: Participation enhances skills in data collection, curation, and AI analytics, positioning practitioners as leaders in their field.
    \item \textbf{Visibility and Recognition}: Contributing to themed challenges boosts practitioners' visibility and reputation within the healthcare community.
    \item \textbf{Community Impact}: Practitioners can improve local healthcare services and address community challenges through AI-driven solutions from curated datasets.
\end{itemize}

\subsection{Strategies for Data Collection and Curation}
The success of this strategic framework relies on robust data collection and curation strategies to ensure dataset integrity, quality, and usability. The proposed strategies include:

\begin{itemize}
    \item \textbf{Academic Projects and Ethical Compliance}: Clinicians can integrate data collection into academic projects, ensuring adherence to ethical standards, patient consent, and compliance with laws and regulations like the Helsinki Declaration, GDPR, and HIPAA.
    \item \textbf{Data Curation Practices}: Follow established practices, such as those from the BraTS challenge, to standardize data formats, document metadata, and maintain quality control.
    \item \textbf{Image Labeling}: Annotate images to highlight normal tissue, lesions, or cancer, providing context for AI-driven supervised learning.
    \item \textbf{Acceptance of Curated Data with and without Labels}: Accept both labeled and unlabeled data to encourage participation from various expertise levels.
\end{itemize}

\subsection{Potential Benefits for Healthcare Outcomes}
Establishing medical image databases across Africa offers significant benefits for healthcare outcomes:

\begin{itemize}
    \item \textbf{Advancement of AI-driven Healthcare Innovation}: Localized datasets will enable the development of AI models tailored to the African context, improving diagnosis, treatment planning, and patient management.
    \item \textbf{Enhanced Resources in Hospitals}: AI-based diagnostic tools will upscale resources in African hospitals, streamlining workflows, optimizing resource allocation, and enhancing efficiency.
    \item \textbf{Empowerment of Practitioners}: Access to AI tools trained on local data empowers healthcare practitioners to make better clinical decisions, improving patient outcomes and care quality.
    \item \textbf{Reduction of Healthcare Disparities}: Democratizing access to AI-driven healthcare reduces disparities, providing underserved communities with timely and effective diagnostic tools and narrowing healthcare outcome gaps.
\end{itemize}
\section{Discussion and Conclusion}
The proposed strategic framework presents a transformative approach to addressing the challenge of neuroimaging data scarcity in African healthcare. Incentivizing healthcare practitioners to establish local databases and curate high-quality datasets through themed challenges can foster a more inclusive and effective ecosystem for AI-driven healthcare innovation..

Healthcare practitioners can contribute their expertise and resources to build a robust repository of locally sourced data, curated to stringent quality and ethical standards. This can serve as the foundation for developing AI models tailored to the unique needs of the African population. The issue of limited local computational resources can be mitigated by collaborating with better-equipped institutions for development and training of AI models using the available datasets. As a future direction, Federated Learning\cite{li2020review} offers a promising method for advancing medical image computation while safeguarding data privacy and security. Clinicians contributing datasets can be introduced to federated learning once their data meet quality standards, enabling them to collaborate on model training.

In conclusion, introduction of themed challenges would be a significant step towards democratizing access to AI-driven healthcare innovation in Africa. By harnessing the power of locally curated data, this strategic framework aims to align problem-solving efforts with long-term research goals rather than short-term solutions. This approach has the potential to initiate a profound change in healthcare delivery, drive innovation, and improve health outcomes for millions of people across the continent.

\clearpage

\bibliographystyle{splncs04} 
\bibliography{ref} 

\end{document}